\documentclass[12pt]{iopart}

\usepackage{graphicx}

\def\eion{{(e~+~ion)}\ }

\def\fexvii{{\rm Fe~\sc xvii}\,}
\def\fexviii{{\rm Fe~\sc xviii}\,}
\def\fexix{{\rm Fe~\sc xix}\,}
\def\fexx{{\rm Fe~\sc xx}\,}

\def\en{{$n$\ }}

\newcommand{\be}{\begin{equation}}
\newcommand{\ee}{\end{equation}}

\begin{document}

\title[R-Matrix calculations for opacities]{R-Matrix calculations for
opacities:~II. Photoionization and oscillator strengths of iron ions 
\fexvii, \fexviii and \fexix}

\author{S N Nahar$^1$, L Zhao$^2$, W Eissner$^3$, A K Pradhan$^1$}

\address{$^1$ Department of Astronomy,
The Ohio State University, Columbus, Ohio 43210, USA}
\address{$^2$ Department of Physics,
The Ohio State University, Columbus, Ohio 43210, USA}
\address{$^3$ Institut f\"ur Theoretische Physik, Teilinstitut 1,
 70550 Stuttgart, Germany\footnote{Deceased}}
\vspace{10pt}

\begin{abstract}
Iron is the dominant heavy element that plays an important role in 
radiation transport in stellar interiors. Owing to its abundance and 
large number of bound levels and transitions, iron ions determine the
 opacity more than any other astrophysically abundant element. 
A few iron ions constitute the abundance and opacity of iron at the 
base of the convection zone (BCZ) at the boundary between the solar 
convection and radiative zones, and are the focus of the present study. 
Together, \fexvii, \fexviii and \fexix contribute 85\% of iron ion 
fractions 20\%, 39\% and 26\% 
respectively, at the BCZ physical conditions of temperature T $\sim$ 
$2.11 \times 10^6$K and electron density N$_e$ = $3.1 \times 10^{22}$cc. 
We report heretofore the most extensive R-matrix atomic calculations 
for these ions 
for bound-bound and bound-free transitions, the two main processes
of radiation absorption. We consider
wavefunction expansions with
218 target or core ion fine structure levels of \fexviii for \fexvii,
276 levels of \fexix \ for \fexviii, in the Breit-Pauli R-matrix (BPRM)
approximation, and
180 LS terms (equivalent to 415 fine structure levels) of
\fexx for \fexix calculations.
These large target expansions which includes core ion excitations to 
n=2,3,4 complexes enable accuracy and convergence of photoionization 
cross sections, as well as inclusion of high lying resonances.
The resulting R-matrix datasets include 
454 bound levels for \fexvii, 1,174 levels for \fexviii, and 1,626 
for \fexix~up to $n\leq$ 10 and $l$=0 - 9. Corresponding datasets of 
oscillator strengths for photoabsorption are: 20,951 transitions
for \fexvii, 
141,869 for \fexviii, and 289,291 for \fexix. Photoionization cross
sections have obtained for all bound fine structure levels 
of \fexvii and \fexviii, and for 900 bound LS states of \fexix.
Selected results demonstrating prominent characteristic
features of photoionization are presented, particularly the
strong Seaton PEC (photoexcitation-of-core) resonances formed 
via high-lying core excitations with $\Delta n=1$ that significantly
impact bound-free opacity. 

\end{abstract}


 \vspace{2pc}

 \submitto{\JPB}

 \maketitle
  
%

\section{Introduction}

 As described in the first paper in this series {\it R-Matrix
calculations for opacities} RMOP1 \cite{petal23},
solar elemental abundances are uncertain which, in turn, are related
to the accuracy of atomic opacities in stellar interiors.
Opacity, which is measure of radiation absorption during its transport, 
is determined by two main processes, absorption by photo-excitations and 
photoionization via all bound states for all ionization stages of all 
elements that exist in the plasma, and hence requires extensive amount 
of atomic data. The present work focuses on high precision atomic data
for these two radiative processes.
Opacity also depends by photon scattering and free-free transitions, 
but their contributions are generally small in most of the frequency
range at high temperatures and densities.

This work is an extension of the Opacity Project (hereafter OP
\cite{op}) which reported findings under the series of {\it Atomic 
data for opacities} 
(hereafter ADOC) papers. We first describe the approximations employed
in the OP and other prior works and their limitations, and
extensions and improvements in the present series.

\subsection{Atomic data calculations for opacities}

Other methods besides the R-matrix method used for large-scale 
calculations of atomic data for opacities are based on
atomic structure calculations for the bound-bound transitions and 
the distorted wave (DW) approximation for photoionization. Under such
approximations,
oscillator strengths and photoionization cross sections are computed 
for all possible bound-bound and bound-free transitions among levels 
specified by electronic configurations included in atomic calculations. 

The DW approximation based on an atomic structure 
calculation couples to the continuum and yields complete and readily
computable datasets for opacities. 
However, since the DW approximation includes only the coupling between 
initial and final states, it is unable to produce autoionizing (AI)
resonances embeded in the continua formed 
from the complex interference between the bound and continuum wavefunction 
expansions involving other levels.  DW models employ the independent 
resonance approximation that treats the bound-bound transition probability 
independently from coupling to the continuum. 
In this paper, we report
developments in the R-matrix calculations with new features that impact
the opacity in contrast to the original OP R-matrix works.

\subsection{Opacity Project R-matrix calculations}

In contrast to atomic structure and DW calculations, 
the R-matrix method accounts inherently for coupling effects due to
electron-electron correlation and introduce autoionizing resonance 
profiles in an {\it ab initio} manner. However, R-matrix calculations
are computationally laborious and entail multiple steps. 
The OP work by M.J. Seaton and collaborators \cite{mjs87,betal87} 
was devoted to the development of the R-matrix method using the close 
coupling (CC) approximation based on implementation framework by P.G. 
Burke and collaborators (e.g. \cite{br75,bt75,pb11}). The R-matrix method
was employed extensively for accurate calculations of radiative data for
energies, oscillator strengths and photoionization cross sections 
systematically for most astrophysically abundant atoms and ions
from hydrogen to iron \cite{op}. The objective of the OP was to determine
the stellar plasma opacity using high accuracy radiative atomic data.
The atomic data under the OP are available through OP database, TOPbase 
\cite{top} and NORAD-Atomic-Data \cite{norad}.

Despite unprecedented effort and advances the OP data are not of
sufficient precision or extent, as revealted with the advent of sophisticated
high resolution observational and experimental set-ups. The original
atomic data from the OP for oscillator strengths ($f$-values) and 
photoionization cross sections ($\sigma_{PI}$) were found to be inadequate 
and of insufficient accuracy, primarily because those data were computed 
in LS coupling without relativistic fine structure effects and with very 
limited configuration interaction. 
While OP data included highly excited states with $n \leq$10, 
the typical angular momentum limit was $l \leq$3. 
Many of these calculations were later repeated for more complete data for 
$l\leq$9 using larger wavefunction expansions, and using the 
BPRM method (these data are available from the NORAD-Atomic-Data database 
\cite{norad}). The main problem for discrepancies was found to be the 
missing physics manifest at high energies. The OP R-matrix work 
reported in the ADOC series used small CC
wavefunction expansion which included a few LS terms of the ground
configuration, or a few configurations of the same n-complex of the 
target or the core ion. Such considerations missed important physical 
effects of Seaton PEC resonances first introduced in \cite{ys87}, and 
Rydberg series resonances corresponding to highly excited
core states (demonstrated extensively 
for oxygen ions \cite{snn98} and subsequent works).
In addition, with high resolution observations fine structure data
were needed in contrast to the OP LS coupling data.

\subsection{Breit-Pauli R-matrix method}

The dynamic package of R-matrix codes has been revised and expanded several 
times.  In the follow-up to OP, the Iron Project \cite{ip}, the R-matrix 
package was extended to include fine structure with relativistic effects 
in the Breit-Pauli approximation (the BPRM method 
\cite{betal95}). Other physical effects such as radiation damping of AI
resonances were also incorporated \cite{znp99,nb3}.
Of particular relevance to this work is one of the latest calculations
on the convergence of 
resonant core ion excitations with increasing $n$ for \fexvii \cite{np16}.

\section{Theoretical framework}

 The BPRM framework is described with particular emphasis on atomic
absorption of radiation in plasma sources.

\subsection{Radiative processes for opacities}

The two main processes are photo-excitation for
a bound-bound transition and photo-ionization for a bound-free transition. 
Photo-excitation of an ion $X^+$,
\begin{equation}
h\nu + X^+ \rightarrow X^{+*}, 
\end{equation}
is related to oscillator strength (or $f$-value) which gives the probability 
of transition. Photoionization can occur directly as
\begin{equation}
h\nu + X^+ \rightarrow e + X^{++}, 
\end{equation}
which is described by a smooth background cross section or through an 
intermediate AI state as
\begin{equation}
h\nu + X^{+} \leftrightarrow (X^+)^{**}  \leftrightarrow e + X^{++}
\end{equation}
which introduces a resonance in the cross section. A doubly excited AI 
state is formed when the photon energy matches that of a Rydberg level,
$E_R = E_c^* - z^2/\nu^2$ where $E_c^*$ is an excited energy of the core
ion, $z$ is the ion charge and $\nu$ is the effective quantum number
with respect to $E_c^*$. The state may autoionize into the continuum, or 
undergo dielectronic recombination if a free electron is captured by 
emission of a photon via radiative decay of the core ion.
The AI resonance can be produced theoretically by including the core 
excitations in the wave function expansion in the close coupling (CC) 
approximation. 

As mentioned above, opacity can also be caused by scattering of photons 
by atoms at all 
frequencies of radiation prevalent in a given environment. However, they are 
much less significant compared to bound-bound and bound-free transitions and 
can be taken care more easily as described in the first paper of the
series RMOP.I.

\subsection{Close coupling approximation and the R-matrix method}

The CC approximation describes the atomic system of (N+1) electrons by a
'target' or the 'core' or the residual ion of N-electrons interacting with
the (N+1)th electron. The total wave function, $\Psi_E$, of the (N+1) electrons
system in a symmetry $SL\pi$ is represented by an expansion as (e.g.
\cite{mjs87})
\begin{equation}
\Psi_E(e+ion) = A \sum_{i} \chi_{i}(ion)\theta_{i} + \sum_{j} c_{j} \Phi_{j},
\end{equation}
where the target ion eigenfunction $\chi_{i}$ is coupled with the
(N+1)$^{th}$
electron function $\theta_{i}$ in a bound or continuum state. 
The summation is over the ground and as many excited
ion states as practical in the CC calculation. 
$A$ is the anti-symmetrization operator. The (N+1)$^{th}$ 
electron with kinetic energy $k_{i}^{2}$ corresponds to
a channel labeled as $S_iL_i\pi_ik_{i}^{2}\ell_i(SL\pi)$, where
$S_iL_i\pi_i$ is the symmetry of the target state $i$. For $k_i^2 < 0$
the channel is closed and the $\Psi_E$ represents a bound state (all
channels closed), and for $k_i^2 > 0$ the channel is open and $\Psi_E$
represents a continuum state. 
In the second sum,
the $\Phi_j$s are bound channel functions of the (N+1)-electron system
that account for short-range electron correlation, and subject to
an orthogonality condition between the continuum
and the bound electron spin-orbital functions. 
Autoionizing resonances arise from 
interference effects among th closed and open channels including 
core ion excitations in the CC wave function expansion.

In the BPRM method \cite{ip,aas11} relativistic effects are included 
in the Breit-Pauli approximation where the Hamiltonian of the 
(N+1)-electron system is
\begin{equation}
H_{N+1}^{\rm BP}= \sum_{i=1}\sp{N+1}\left\{-\nabla_i\sp 2 - \frac{2Z}{r_i}
 + \sum_{j>i}\sp{N+1} \frac{2}{r_{ij}}\right\} +
 H_{N+1}^{\rm mass} + H_{N+1}^{\rm Dar} + H_{N+1}^{\rm so}
 \end{equation}
The curly bracketed term is the non-relativistic Hamiltonian and the
additional terms are the relativistic one-body correction terms, the mass
correction, $H^{\rm mass} = -{\alpha^2\over 4}\sum_i{p_i^4}$, Darwin,
$H^{\rm Dar} = {Z\alpha^2 \over 4} \sum_i{\nabla^2({1 \over r_i})}$, and
the spin-orbit interaction, $H^{\rm so}= Z\alpha^2 \sum_i{1\over r_i^3} 
{\bf l_i.s_i}$ where $p_i$ is the momentum of an electron, $\alpha$ is the
fine structure constant, and ${\bf l_i,s_i}$ are the orbital and spin angular

Substitution of the CC wavefunction $\Psi_E(e+ion)$ in the Schrodinger equation
\begin{equation}
H_{N+1}\mit\Psi_E = E\mit\Psi_E
\end{equation}
results in a set of coupled equations that are solved using the R-matrix
method 
\cite{pb11,mjs87,betal95,betal95,aas11}. 
The BPRM method implements an intermediate coupling scheme. The set
of ${SL\pi}$ terms is recoupled for $SLJ\pi$ fine structure
levels of the \eion system, 
followed by diagonalization of the (N+1)-electron BP Hamiltonian. 
The solutions are either a continuum
wavefunction $\Psi_F$ for an electron with positive energies (E $\geq$
0), or
a bound state wavefunction $\Psi_B$ for negative total energies (E $< 0$).

\subsection{R-matrix method for radiative data}

With wavefunction expansions in the R-matrix formulation as above,
 transition matrix element for a radiative transition to an excited
bound state
or for photoionization is given by (e.g. \cite{aas11})
\begin{equation}
<\Psi_j|| {\bf D}|| \Psi_{k}>,
\end{equation}
where the photon-ion interaction is represented by the dipole operator,
${\bf D} = \sum_i{{\bf r}_i}$, and the sum is over the number of active
electrons.

The generalized line strength ${\bf S}$ is expressed as
\begin{equation}
{\bf S}= |<\Psi_j||{\bf D}_L||\Psi_k>|^2 =
 \left|\left\langle{\mit\psi}_f
 \vert\sum_{j=1}^{N+1} r_j\vert
 {\mit\psi}_i\right\rangle\right|^2 \label{eq:SLe},
\end{equation}
where $\mit\Psi_j$ and $\mit\Psi_k$ are initial and final state
wavefunctions. 
The line strengths are energy independent quantities. The
oscillator strengths ($f_{ij}$) and radiative decay rates or Einstein
A-coefficients for E1 dipole transitions are given by

\begin{equation}
f_{ij} = {E_{ji}\over {3g_i}}S(ij), ~~
A_{ji}(a.u.) = {1\over 2}\alpha^3{g_i\over g_j}E_{ji}^2f_{ij}.
\end{equation}
 $E_{ji}$ is the energy difference between initial and final states,
and $g_i$, $g_j$ are
statistical weight factors, respectively.

The photoionization cross section ($\sigma_{PI}$) is obtained as,
\begin{equation}
\sigma_{PI} = {4\pi^2 \over 3c}{1\over g_i}\omega{\bf S},
\end{equation}
 where $g_i$ is the statistical weight factor of the initial
bound state and $\omega$
is the incident photon energy. 
The complex resonant structures in photoionization result from channel
 couplings between open continuum channels ($k_i^2~\geq$ 0) and
closed channels ($k_i^2~<$ 0) at electron energies $k_i^2$ corresponding
to autoionizing states along Rydberg series $S_iL_i~J_i\pi_i\nu_i
\ell_i$,
where $\nu_i$ is the effective quantum number relative to the target
threshold $S_iL_i~J_i\pi_i$.  We note that 
present work also includes radiation damping of the resonances using
the approach of \cite{znp99}, but is found to be insignificant for the
Fe ions considered herein.

\section{R-matrix Computations for atomic processes}

The relativistic R-matrix calculations are carried out through a package 
of BPRM codes \cite{betal95,np94,znp99} in several stages of computations 
as shown in Figure~1 (left branch).
 \begin{figure}
\begin{center}
\hskip -0.35in
 \includegraphics[height=3.7in,width=5.5in]{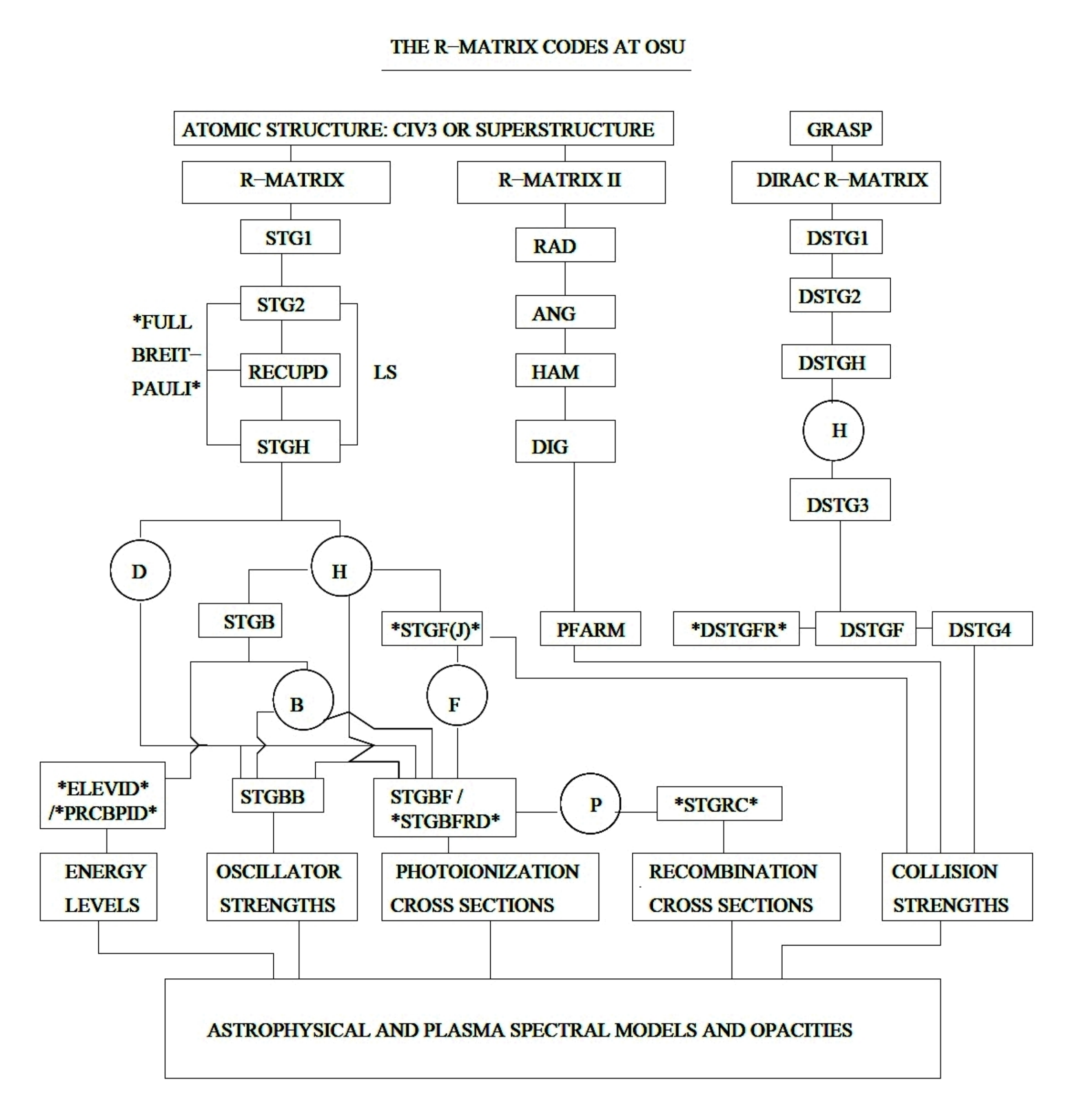}
\end{center}
 \caption{
Schematic diagram for various stages of R-matrix computations. The data 
obtained are: 1) energy Levels, 2) oscillator
strengths, 3) photoionization cross sections, 4) recombination rate
 coefficients, and 5) electron impact collision strengths, 
astrophysical models and opacities calculations.
\label{fig:f1}
}
 \end{figure}

BPRM computations begin with the STG1 program for which the program 
reads the orbital wavefunctions and potentials of the core 
ion as the first input and computes one- and two-electron radial
integrals for the output.
In the present cases for \fexvii, \fexviii and \fexix , these 
wavefunctions are obtained from configuration interaction atomic 
structure calculations using the code {\footnotesize SUPERSTRUCTURE} 
(SS) \cite{ss74,ss03}.
The second program STG2 computes spin-angular algebraic coefficients 
in LS coupling for the ion and \eion Hamiltonian matrices. 
Program RECUPD recouples the $LS$ coefficients in intermediate coupling
to introduce fine structure $SLJ$. Using SS 
wavefunctions, RECUPD also recomputes fine structure energies of the
core ion. Typically the energy values and order from 
RECUPD match closely with those from SS. However, for complex ions
they may differ in the third or fourth significant digits and energy order for
some levels, particularly those highly excited ones. The program STGH 
diagonalizes the \eion 
Hamiltonian to obtain R-matrix basis functions that are used to
compute subsequent parameters as follows.
Program STGB computes bound energy 
levels and wavefunctions and STGF computes continuum wavefunctions and 
electron impact excitation collision strengths. STGBB computes oscillator 
strengths for bound-found transitions, and STGBF computes
photoionization cross sections for bound-free transitions.

\subsection{Core ion wavefunctions from SUPERSTRUCTURE}

As mentioned above, computation using R-matrix codes starts with 
wavefunctions of the core ion obtained from code SUPERSTRUCTURE (SS). 
Similar to intermediate coupling in the BPRM method, 
SS includes relativistic contributions in Breit-Pauli 
approximation \cite{ss03}. SS includes several terms of the Breit
interaction in addition to one-body terms in BPRM calculations, such
as the full Breit interaction and part of other two-body interactions. 
Configuration interaction calculations are carried out using
the Thomas-Fermi-Dirac-Amaldi central potential to compute one-orbitals 
functions, scaled according to a variational minimization scheme
\cite{ss74,ss03,aas11}.

The list of configurations and Thomas-Fermi orbital scaling parameters 
for core ions of each of the three Fe ions \fexvii, \fexviii, \fexix~
are given and discussed
in following subsections. All configurations for each ion are treated as 
spectroscopic; that is, all energies are optimized in SS iteratively. 
Each table quotes 
the total number of core ion excitation produced by the spectroscopic 
configurations, all of which were included in wavefunction expansions. 

The SS run itself computes a large number of the transitions of 
types E1, E2, E3, 
M1, and M2 among all possible levels possible within configurations
specified in atomic structure calculations. Although
not presented in this paper which focuses on R-matrix results for
E1 transitions that primarily contribute to opacities, SS results for
all energy levels and all types of transitions stated here 
are available through atomic database NORAD-Atomic-Data \cite{norad}.

Progran RECUPD of the R-matrix codes use orbital wavefunctions
obtained from SS to recompute the energies of the core ion. These 
reproduced energies match almost exactly to those from SS for most 
ions. However, with large complex atomic systems the SS and RECUPD 
energies can show differences in the third or fourth decimal figures, 
and also the order of the higher energy levels. In the present work,
for both \fexvii~ and \fexviii, the RECUPD energies are used in the 
Hamiltonian matrix diagonalization in STGH.

\subsubsection {CC wavefunction expansion for \fexvii:}

Table~\ref{tab:tab1} lists the optimized set of 17 configurations with 
Thomas-Fermi scaling parameters of orbitals that produced the 218 energy
evels for the core ion \fexviii~ included in the first summation term 
on RHS of Eq.~(4) to represent the \eion wavefunction expansion for \fexvii.
Table~\ref{tab:tab1} provides only a small sample set of energies of the 
ground and a few excited levels of \fexviii~ from the full set of 218 
levels obtained from SS.  SS energies are compared with 
measured values by \cite{metal85} available from
NIST \cite{nist}, and found to be in good agreement.

\begin{table}
\caption{
Optimized set of 17 spectroscopic configurations of the core ion
\fexviii belonging to complexes n=2,3,4 are:
$1s^22s^22p^5$(1), $1s^22s2p^6$(2), $1s^22s^22p^43s$(3), $1s^22s^22p^43p$(4),
$1s^22s^22p^43d$(5), $1s^22s2p^53s$(6), $1s^22s2p^53p$(7),
$1s^22s2p^53d$(8), $1s^22s^22p^44s$(9), $1s^22s^22p^44p$(10),
$1s^22s^22p^44d$(11), $1s^22p^63s$(12), $1s^22p^63p$(13), $1s^22p^63d$(14),
$1s^22s2p^54s$(15), $1s^22s2p^54p$(16), $1s^22s2p^54d$(17).
Thomas-Fermi orbital scaling parameters are: 1.39944(1s), 
1.20409(2s), 1.14074(2p), 1.13092(3s), 1.08125(3p), 1.11030(3d), 
1.13092(4s), 1.08125(4p), 1.1103(4d). 
The table present a small sample set of energies from the 218-level set, 
being compared with those of \cite{metal85} available at NIST webpage 
\cite{nist}.
 The number within parentheses next to the LS term corresponds to 
configuration number for the term.
\label{tab:tab1}
}
 \footnotesize
\begin{tabular}{llcllllcll}
\br
 \multicolumn{1}{r}{i} & \multicolumn{1}{c}{LS term}&
 \multicolumn{1}{c}{2J+1} & \multicolumn{1}{c}{E$_T$(Ry)} &
 \multicolumn{1}{c}{E(Ry)} &
 \multicolumn{1}{l}{i} & \multicolumn{1}{c}{LS term} &
 \multicolumn{1}{c}{2J+1} & \multicolumn{1}{c}{E$_T$(Ry)}&
 \multicolumn{1}{c}{E(Ry)} \\
\mr
 \multicolumn{10}{c}{Total number of levels = 218}\\
\mr
   1& $^2P^o$(1)&   4&  0.0        & 0.0 &
   2& $^2P^o$(1)&   2&  0.94116     & 0.9348 \\
   3& $^2S^e$(2)&   2&  9.8248     & 9.70 & 
   4& $^4P^e$(3)&   6&  56.798     & 57.70 \\
   5& $^2P^e$(3)&   4&  57.054     & 56.94 &
   6& $^4P^e$(3)&   2&  57.502     & 57.50 \\
   7& $^4P^e$(3)&   4&  57.668     & 57.57 &
   8& $^2P^e$(3)&   2&  57.906     & 57.80 \\
   9& $^2D^e$(3)&   6&  58.436     & 58.32 & 
  10& $^2D^e$(3)&   4&  58.471     & 58.36 \\
  \multicolumn{8}{c}{...}\\
\br
 \end{tabular}
 \end{table}
The Hamiltonian matrix for \fexvii is set up and diagonalized in STGH using 
these energies and energy order of the core ion reproduced in RECUPD.
Partial waves 0$\leq\ell\leq$ 9 for the interacting free electron form
singlet, triplet, and quintet spins for \eion $LS\pi$
up to $L$=0-4 of even and odd parities, 
recoupled in RECUPD to yield total $SLJ\pi$ symmetries with $J \leq 12$.
The R-matrix boundary was chosen to be $a_o = 6$ a.u., 
large enough to ensure all bound orbitals to have decayed to at least
$P_{n\ell} < 10^{-3}$. 

The second term in RHS of Eq.~(4), which represents \eion bound state
correlation functions as bound channels in the Hamiltonian, 
included 42 ($N$+1)-particle configurations with 
minimum to a maximum number electron occupancies in orbitals as given 
within parentheses: 1s(2-2), 2s(0-2), 2p(3-6), 3s(0-2), 
3p(0-2), 3d(0-2), 4s(0-1), 4p(0-1), 4d(0-1). The total angular momenta
of \fexvii~selected are J=0-9 of even and odd parities, sufficient for 
radiative excitations via dipole E1 transitions.

\subsubsection{CC wavefunction expansion for \fexviii:}

The ground and 275 excited fine structure levels of the core ion \fexix~ 
included with configuration complexes $n$=2,3,4, were
optimized using a set of 12 configurations given in Table~\ref{tab:tab2}
along with the Thomas-Fermi scaling parameters of orbitals, and a small
subset of the 276-level expansion for \fexviii.
The calculated energies from SS are compared with measured values
from NIST \cite{nist}.
\begin{table}
\caption{
The set of 12 spectroscopic configurations of the core ion \fexix~ (as in
table \ref{tab:tab1}), optimized for energies and orbitals with
complexes n=2,3,4 in the CC wavefunction expansion of \fexviii~ are: 
$1s^22s^22p^4$(1), $1s^22s2p^5$(2), $1s^22p^6$(3), $1s^22s^22p^33s$(4), 
$1s^22s^22p^33p$(5), $1s^22s^22p^33d$(6), $1s^22s^22p^54s$(7), 
$1s^22s^22p^34p$(8), $1s^22s^22p^34d$(9),
$1s^22s2p^43s$(10), $1s^22s2p^43p$(11), $1s^22s2p^43d$(12). 
The set of Thomas-Fermi scaling parameters for the orbitals are 
1.35(1s), 0.9009(2s), 1.12(2p), 1.07(3s), 1.05(3p), 1.10(3d), 1.0(4s), 
1.0(4p), 1.0(4d). 
\label{tab:tab2}
}
 \footnotesize
\begin{tabular}{llcllllcll}
\br
 \multicolumn{1}{r}{i} & \multicolumn{1}{c}{LS term}&
 \multicolumn{1}{c}{2J+1} & \multicolumn{1}{c}{E$_T$(Ry)} &
 \multicolumn{1}{c}{E(Ry)} &
 \multicolumn{1}{l}{i} & \multicolumn{1}{c}{LS term} &
 \multicolumn{1}{c}{2J+1} & \multicolumn{1}{c}{E$_T$(Ry)}&
 \multicolumn{1}{c}{E(Ry)} \\
\mr
 \multicolumn{10}{c}{Total number of levels = 276}\\
\mr
   1& $^3P^e$(1)&   4&  0.0       & 0.0 & 
   2& $^3P^e$(1)&   0&  0.6681    &  0.6857 \\
   3& $^3P^e$(1)&   2&  0.8003    &  0.8150 & 
   4& $^1D^e$(1)&   4&  1.5675    & 1.5387  \\
   5& $^1S^e$(1)&   0&  2.9543    & 2.9629 & 
   6& $^3P^o$(2)&   4&  8.4592    & 8.4100 \\
   7& $^3P^o$(2)&   2&  9.0267    & 8.9736 & 
   8& $^3P^o$(2)&   0&  9.4316    & 9.3862 \\
   9& $^1P^o$(2)&   2&  11.714    & 11.5512 & 
  10& $^1S^e$(3)&   0&  19.706    & 19.4481 \\
  \multicolumn{8}{c}{...}\\
%
 \br
 \end{tabular}
 \end{table}

The CC expansion for \fexviii~ included  0$\leq\ell\leq$ 9 partial waves 
with doublet, quartet, sextet
$LS$ symmetries and $L$=0-4 of even and odd parities of the core ion \fexix. 
The R-matrix basis sets for orbitals contained 20 continuum functions
inside the R-matrix boundary of radius 4 $a_o$. 
The second bound-channel correlation function term in Eq.~(4) included 
96 ($N$+1)-electron configurations with minimum to a maximum occupancies 
in orbital as given within parentheses: 1s(2-2), 2s(0-2), 2p(2-6), 3s(0-2), 
3p(0-2), 3d(0-2), 4s(0-2), 4p(0-2), 4d(0-1). The total \eion~ angular 
momenta $SLJ\pi$ for \fexviii~ was chosen to be $J$=1/2-17/2 of even and 
odd parities.

\subsubsection{CC wavefunction expansion for \fexix:}

The atomic data for transition probabilities with fine structure
of \fexix~ was computed using BPRM codes.
Line and oscillator strengths for bound-bound transitions were
computed an 18CC wavefunction expansion for the core ion \fexx~
(details in \cite{snnfe19}) since no bound states of \fexix are formed 
with higher core excitations. 
The computation of photoionization cross sections was initially
set with a very large 415CC wavefunction expansion which included n=2,3,4
core ion excitations 
and resonances up to the high energy region. 
However, owing to computational limits BPRM computations proved to 
be impractical.  Hence $\sigma_{PI}$ were obtained in the LS coupling 
approximation.  Table~\ref{tab:tab3} lists the optimized set of 
configurations with Thomas-Fermi scaling parameters of orbitals of \fexx,
and provides a sample set of energies for the ground and a number of 
excited states, compared with energies from NIST \cite{nist}. A large 
number of odd parity states exist in the high energy region, but they are 
not allowed for dipole transitions from the ground state $2s^22p^3(^4S^o)$
and therefore corresponding series of strong 
resonances would not manifest themselves. Hence,
a concise set of 56 LS states was chosen which includes
all dipole allowed and forbidden transitions from the ground and low-lying 
states, where the basic physics of transitions with the full set of 415 fine
structure levels is retained. 
\begin{table}
\caption{
The set of 16 spectroscopic configurations of the core ion \fexx~ that 
was optimized for \fexix~ CC calculations with complexes n=2,3,4 are:
$2s^22p^3$(1), $2s2p^4$(2), $2p^5$(3), $2s^22p^23s$(4), $2s^22p^23p$(5), 
$2s^22p^23d$(6), $2s^22p^24s$(7), $2s^22p^24p$(8), $2s^22p^24d$(9), 
$2s^22p^24f$(10), $2s2p^33s$(11), $2s2p^33p$(12), $2s2p^33d$(13), 
$2s2p^34s$(14), $2s2p^34p$(15), $2s2p^34d$(16) with filled $1s$ orbital.
Thomas-Fermi scaling parameters are:
1.35(1s),  1.25(2s), 1.12(2p), 1.07(3s), 1.05(3p), 1.0(3d), 1.0(4s), 
1.0(4p), 1.0(4d), 1.0(4f).
A sample set of energies $E_T$ is compared
with $E$ \cite{metal85} available from NIST \cite{nist}. 
\label{tab:tab3}
}
 \footnotesize
\begin{tabular}{llllllll}
\br
 \multicolumn{1}{r}{i} & \multicolumn{1}{c}{LS term}&
  \multicolumn{1}{c}{E$_T$(Ry)} &
 \multicolumn{1}{c}{E(Ry)} &
 \multicolumn{1}{l}{i} & \multicolumn{1}{c}{LS term} &
 \multicolumn{1}{c}{E$_T$(Ry)}&
 \multicolumn{1}{c}{E(Ry)} \\
\mr
 \multicolumn{8}{l}{Total number of levels = 415 and states =180}\\
 \mr
 \noalign{\smallskip}
  1& $2s^22p^3(       ^4S^o)$& 0.0       & 0.0 & 
  2& $2s^22p^3(       ^2D^o)$& 1.5091  & 1.4683  \\ 
  3& $2s^22p^3(       ^2P^o)$& 2.7698  & 2.7549 & 
  4& $2s2p^4(         ^4P )$& 7.2235  & 7.2029 \\
  5& $2s2p^4(         ^2D )$& 9.6995  & 9.5869 & 
  6& $2s2p^4(         ^2S )$&10.9879  &10.8920 \\ 
  7& $2s2p^4(         ^2P )$&11.7840  &11.6184 &  
  8& $2p^5(           ^2P^o)$&18.3096  &18.1380 \\
  9& $2s^22p^23s(     ^4P )$&66.4611  &66.2742 & 
 10& $2s^22p^23s(     ^2P )$&66.6114  &  \\
\br
 \end{tabular}
 \end{table}

The LS term energies computed in STG2 are in similar order as that os SS 
but with slight differences in values in the 3rd or 4th figure and 
a few energies with different order. 
Partial waves for the \fexix~ calculations included 0$\leq\ell\leq$ 9 
and formed 89 singlets, triplets, quintets and septets, with the target 
ion total 
$L$=0-4 of even and odd parities. The R-matrix basis sets contained 14 
continuum wavefunctions, R-matrix boundary was $a_o$ = 6 a.u.  The bound 
channel correlation functions included 104 ($N$+1)-particle 
configurations with a minimum to a maximum number electron occupancies in the 
orbitals as given within parentheses: 1s(2-2), 2s(0-2), 
2p(2-6), 3s(0-2), 3p(0-2), 3d(0-2), 4s(0-1), 4p(0-1), 4d(0-1), 4f(0-1).

\subsection{Bound states and oscillator strengths}

The bound energies were obtained by numerically scanning through eigenvalues 
of the \eion Hamiltonian with a sufficiently fine energy mesh in 
effective quantum number, 
typically 0.001-0.005, and corresponding wavefunctions were computed
using program STGB of the R-matrix codes (Figure~\ref{fig:f1}).
Comparisons show good agreement between the observed NIST compilation and
calculated energies \cite{ss03,snnfe18,snnfe19}. 
It may be noted that the R-matrix calculations encompass a large 
number of configurations for the (N+1)-electrons atomic system, and
generally more accurate and yield more extensive data sets than atomic
structure codes such as SS.

The transition parameters, such as, oscillator strengths, line strengths,
and radiative decay rates for the iron ions \fexvii-\fexix
\cite{ss03,snnfe18,snnfe19} were obtained using 
program STGBB of the R-matrix codes which uses the Hamiltonian matrix and
dipole transition matrices computed by STGH and bound wavefunctions
computed by STGB.

\subsection{Bound-free photoionization cross sections}

Basic physical features and illustrative results from large-scale
computations of photoionization cross sections ($\sigma_{PI}$)
of the three Fe ions \fexvii, \fexviii and \fexix~ were obtained using
the STGBF program of the R-matrix package of codes. The $\sigma_{PI}$ 
of \fexvii~ and \fexviii~ reported herein have been obtained from new 
BPRM calculations. Whereas the $\sigma_{PI}$ of \fexix~ are taken from 
\cite{snnfe19px} but features relevant to opacities calculations are 
highlighted and discussed for comparison, consistency and completeness.

Owing to large sizes of arrays and matrices, the BPRM codes went 
through an extensive revisions for opacities calculations.
Often the computations could be carried out only for few energy levels and
photon energies
at a time, and required several years to complete. 

Photoionization cross sections are obtained with consideration of radiation
damping implemented in BPRM codes
\cite{betal95,np94,znp99}, although not eventually found to be
significant for opacities for these Fe ions. 
Autoionizing resonances in photoionization span wide energy ranges,
and were resolved with variable and appropriately fine energy meshes.

\section{Results and discussion}

Opacity calculations require complete datasets for any and all atomic
species.
We report more extensive studies of the three Fe ions, \fexvii, \fexviii 
and \fexix, than previous works and which also reveal
reveal new features in photoionization not 
observed before. With the objective of obtaining high accuracy and complete
sets of atomic data we calcauted
transition probablities and photoionizatio cross sections of
levels with $n \leq 10$ and $l \leq 9$, and all associated $SLJ\pi$ 
spin-orbital symmetries, with 
large wavefunction expansions that show important features in
high energy regions. 
Selected results and prominent characteristic features are discussed below.

\subsection{Energy levels and Oscillator strengths}

We obtained large sets of fine structure energy levels for the three
Fe ions \fexvii, \fexviii~ and \fexix~ \cite{ss03,snnfe18,snnfe19}.
The number of  energy levels and oscillator strengths obtained from 
BPRM method for each ion are given
in Table~\ref{tab:tab4}. For calculating oscillator strengths between
bound-bound transitions, size of the CC wavefunction expansions was 
smaller compared to those for photoionizatio cross sections, since 
very high excited core ion levels do not lead to additional bound levels 
of the \eion system. However, the larger number of core ion thresholds
included herein for photoionization give rise to many more high lying
Rydberg series of autoionizing states that can be much stronger than
those from the low lying excited states.

\begin{table}
\caption{The table lists the number of energy levels and oscillator 
strengths for bound-bound transitions obtained for three iron ions.
The numbers are nearly the same as those obtained in earlier works with
much smaller wavefunction expansions. 
\label{tab:tab4}
}
 \footnotesize
\begin{tabular}{cccl}
\br
 \multicolumn{1}{c}{Ion} & \multicolumn{1}{r}{No. of energies} &
 \multicolumn{1}{r}{No. of oscillator strengths} & \multicolumn{1}{l}{Reference} \\
\mr
 Fe~XVII &   454  &  20,951 &  Nahar et al (2003)\cite{ss03} \\
 Fe~XVIII & 1174  & 141,869 &  Nahar  (2006)\cite{snnfe18} \\
 Fe~XIX   & 1626  & 289,291 &  Nahar (2011)\cite{snnfe19}\\
\br
 \end{tabular}
 \end{table}

\subsection{Photoionization cross sections}

The BPRM method reveals several characteristic features in
photoionization that are of importance in opacity calculations. The 
features can be characterized based on type of states or levels of 
the particular ion, and can impact opacities differently depending 
on energy, temperature and density of plasma in a given region. 
The following subsections focus on these features.

Resonances in photoionization may play a dominant role as they
can increase radiation absorption by orders of magnitude.  In particular,
the present work shows that one of the main reasons for discrepancy 
in photoabsorption from past studies is the lack 
of consideration of resonances due to high lying core excitations. 
The only way to obtain 
these resonances inherently is through the close coupling approximation 
from a large wavefunction expansion 
that includes sufficiently high excitations.
Hence, we include many excited levels belonging 
to n=2, 3 and 4 complexes for the 3 Fe ions to satisfy this convergence
criterion (see also RMOP.III).

\subsubsection{Photoionization of ground states}

\begin{figure}
  \includegraphics[height=4.5in,width=5.0in]{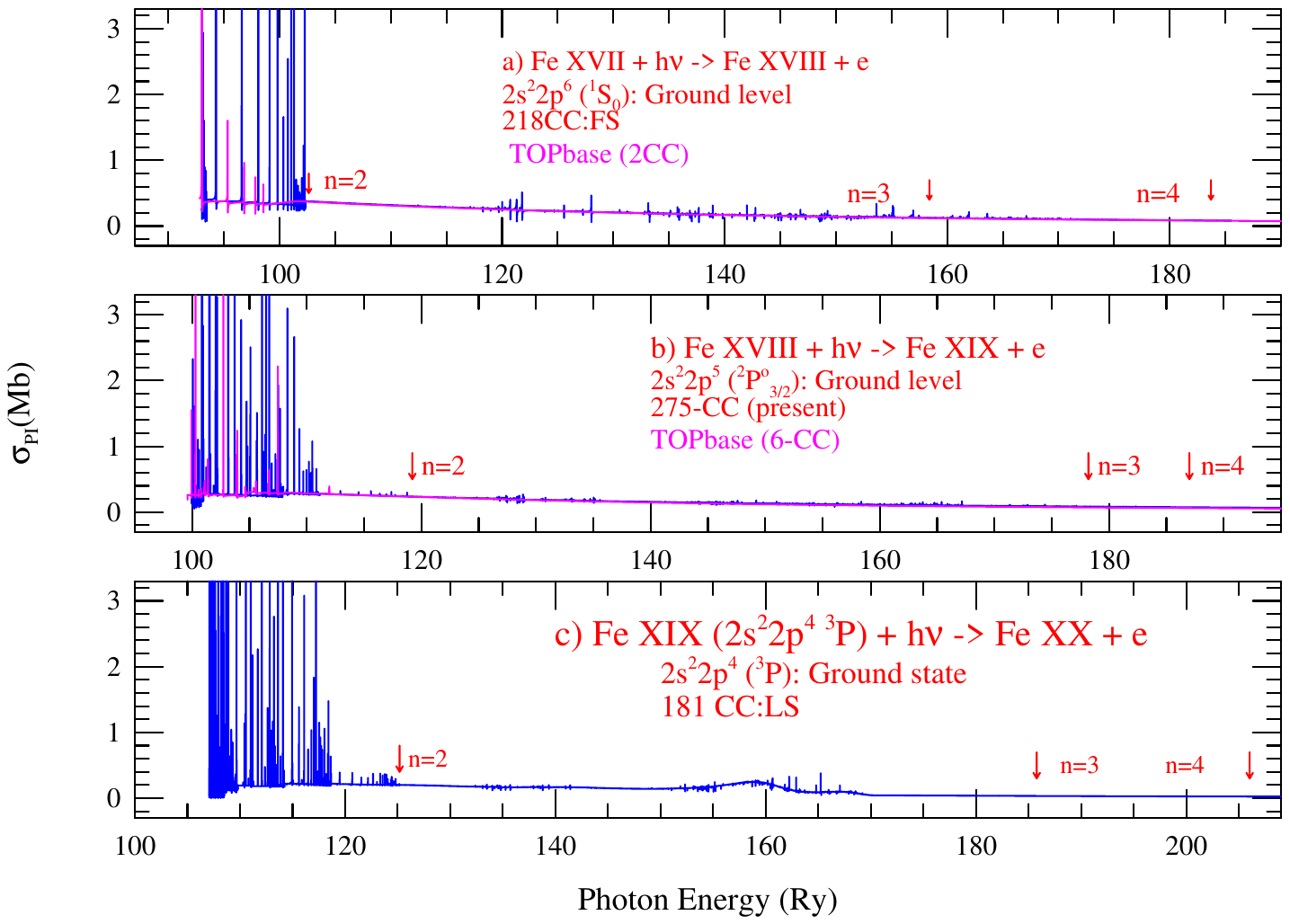}
 \caption{
$\sigma_{PI}$ of the ground level of a) Fe~XVII (218CC FS), 
b) Fe~XVIII (276CC FS), c) Fe XIX (180CC LS).
Present computations with large wavefunction expansion for $\sigma_{PI}$ 
in blue basically agree with those in magenta from earlier TOPbase
calculations with only the ground \en=2 configuration CC expansion \cite{top}.
Hence, photoionization of the ground level is largely unaffected by
inclusion of high-level excitations.
\label{fig:f2}
}
 \end{figure}

The accuracy of the ground level and associated transitions are
obviously important in all applications. 
However, in a plasma at different temperatures and densities 
the photoionization cross section $\sigma_{PI}$ of the ground state 
is not necessarily the most significant one, and in fact may not
dominate bound-free opacity \cite{nb4}. Typically, $\sigma_{PI}$
has a slowly varying background cross section up to the highest
threshold energy in the CC expansion, and then decreases with energy.
The Rydberg series of resonances are superimposed up on the background. 
However, it is the \en-complex of the core ion ground and low-lying 
configurations that produce more prominent resonances compared to 
higher ones, as their magnitudes weaken.
Figure~\ref{fig:f2} 
presents photoionization cross sections of the ground states of
the three ions Fe~XVII-XIX from the present work (blue) and from TOPbase
\cite{top} (magenta) for Fe~XVII and Fe~XVIII. No other detailed
$\sigma_{PI}$ for Fe~XIX are found in literature. 
Our study finds that ground state core excitations beyond
\en=2 complex are not important as they do not produce strong
resonances, and inclusion of \en = 3 and 4 levels does
 not impact cross sections significantly.

\subsubsection{Relativistic fine structure effect:}

In LS coupling resonances in $\sigma_{PI}$ are approximately averaged 
 over their fine structure components. Accuracy increases with 
 inclusion of relativistic fine structure channels as they provide more 
 resolved resonances, more accurate positions of resonances spread over 
 somewhat more extended energy region, as well as additional resonances 
 not allowed in LS coupling.
 With splitting of LS terms of the core ion states into fine 
 structure levels, a much larger number of excited core ion thresholds 
is produced, resulting in correspondingly larger number of Rydberg 
 series of resonances.
 However, the resulting accuracy in $\sigma_{PI}$ may not be significant 
 when the 
 resonances are statistically averaged to obtain integrated quantities 
 such as recombination rates or photoionization rates for plasma opacity 
 at high temperature-density. 
But exceptions are noticeable at low energies and in low 
 temperature plasmas when fine structure coupling creates resonant features, 
 which are actually observed in experiments (e.g. \cite{snn02,ketal99}), 
 but missing from LS coupling calculations. 

\begin{figure}
 \vskip -0.25in
  \includegraphics[width=5.5in,height=4.5in]{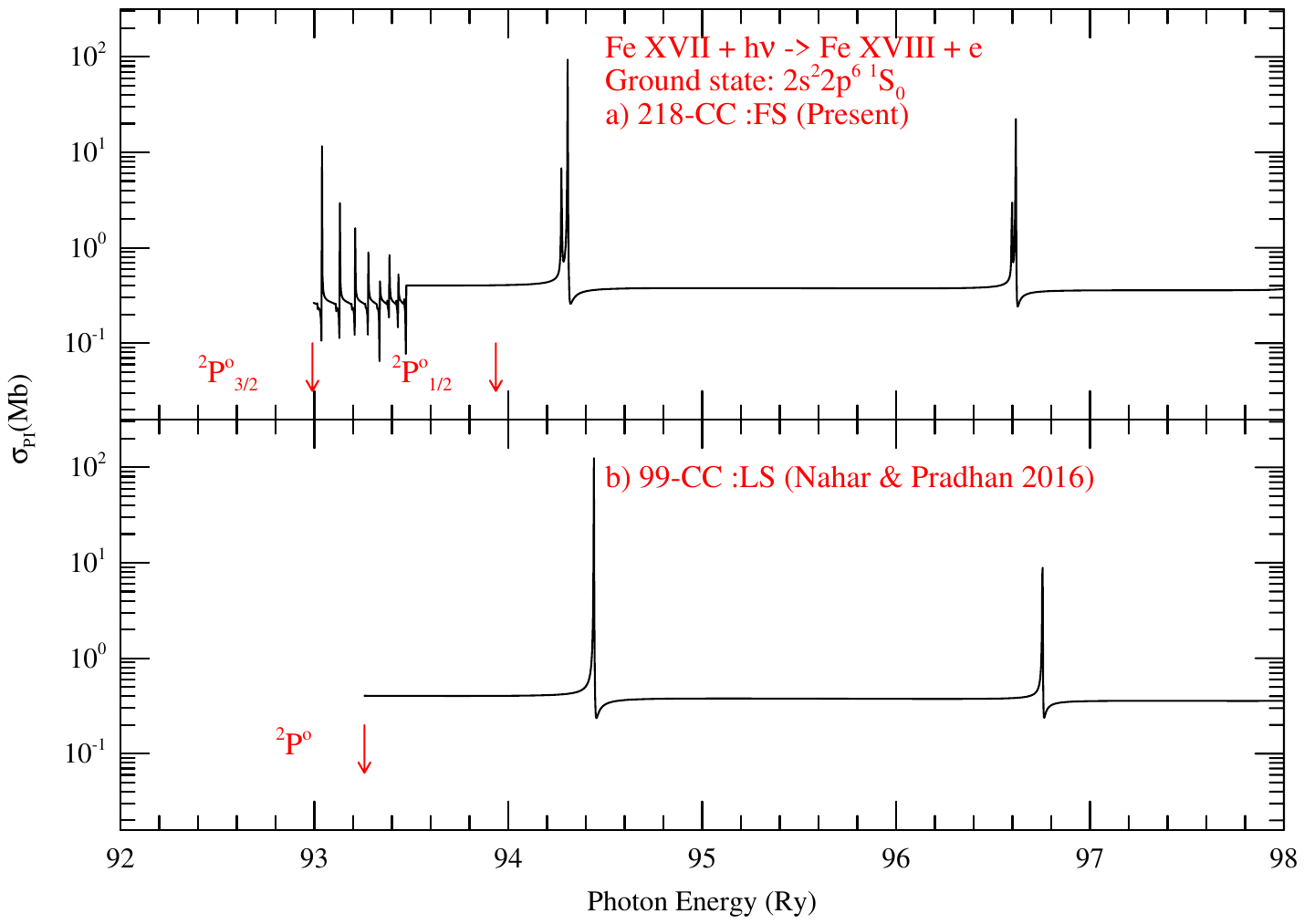}
 \caption{Photoionization cross sections $\sigma_{PI}$ in BPRM 
 method of the ground level $2s^22p^6(^1S_{0})$ of Fe~XVII in the 
 near ionization threshold region (upper panel), and non-relativistic 
 LS coupling R-matrix cross sections (lower panel, \cite{np16}), showing
 the coupling effect of 
 fine structure channels in the region between $2s^22p^5(^2P^o_{3/2})$ 
 and $2s^22p^5(^2P^o_{1/2})$ in the upper panel with resonances 
 and background jump at $2s^22p^5(^2P^o_{1/2})$, not
 formed in LS coupling.
\label{fig:f3}
 } 
      \end{figure}

 Figure~\ref{fig:f3} demonstrates the effect of coupling of relativistic fine 
 structure channels on photoionization in the near ionization energy for 
 the ground $2s^22p^6(^1S_0)$ state of \fexvii. The upper panel is the 
 present BPRM $\sigma_{PI}$,
 and the lower panel from a non-relativistic R-matrix calculation in
 LS coupling \cite{np16}. Features in both cross sections are very similar.
 However, the upper panel shows resonances created by fine structure
 channels $2s^22p^5(^2P^o_{1/2})\epsilon s,d$ in the energy region between
 the two ground state core ion levels $2s^22p^5(^2P^o_{3/2})$ 
and $2s^22p^5(^2P^o_{1/2}$,
 and an enhancement at the $2s^22p^5(^2P^o_{1/2})$ threshold, unlike in
 coupling without fine structure splitting of $2s^22p^5(^2P^o)$.
 Also, the ionization threshold is lowered to the 
$2s^22p^5(^2P^o_{3/2})$ (pointed arrow), whereas in LS coupling the 
 threshold is at the statistical average of the two levels. 
 These resonances would affect applications
 in low energy-temperature plasma sources. 
 In addition, it may be noted that fine structure has split the 
 resonances in LS coupling in lower panel into its component resonances 
 in the upper paner.
 It has been found for the ion that relativistic fine structure effect
 near the ionoization threshold exists in $\sigma_{PI}$ of most of 
 the excited levels of the ion.

\subsubsection{Photoionization of equivalent electron states:}

Equivalent electron levels/states, with more than a single electron 
in the outer orbit, typically have photoionization cross sections 
with smooth background with some enhancement at core ion thresholds,
and then decrease slowly with energy. These levels, particularly the ones 
formed from excited configurations, produce high-peaked closely-spaced
Rydberg series of resonances at lower energies. 
These resonances typically belong to the core ion excitations of the 
same \en-complex as the ground state. 
Hence these states can have significant impact on applications in 
relatively low temperature plasmas. 

\begin{figure}
\hskip -0.35in
  \includegraphics[height=3.7in,width=5.5in]{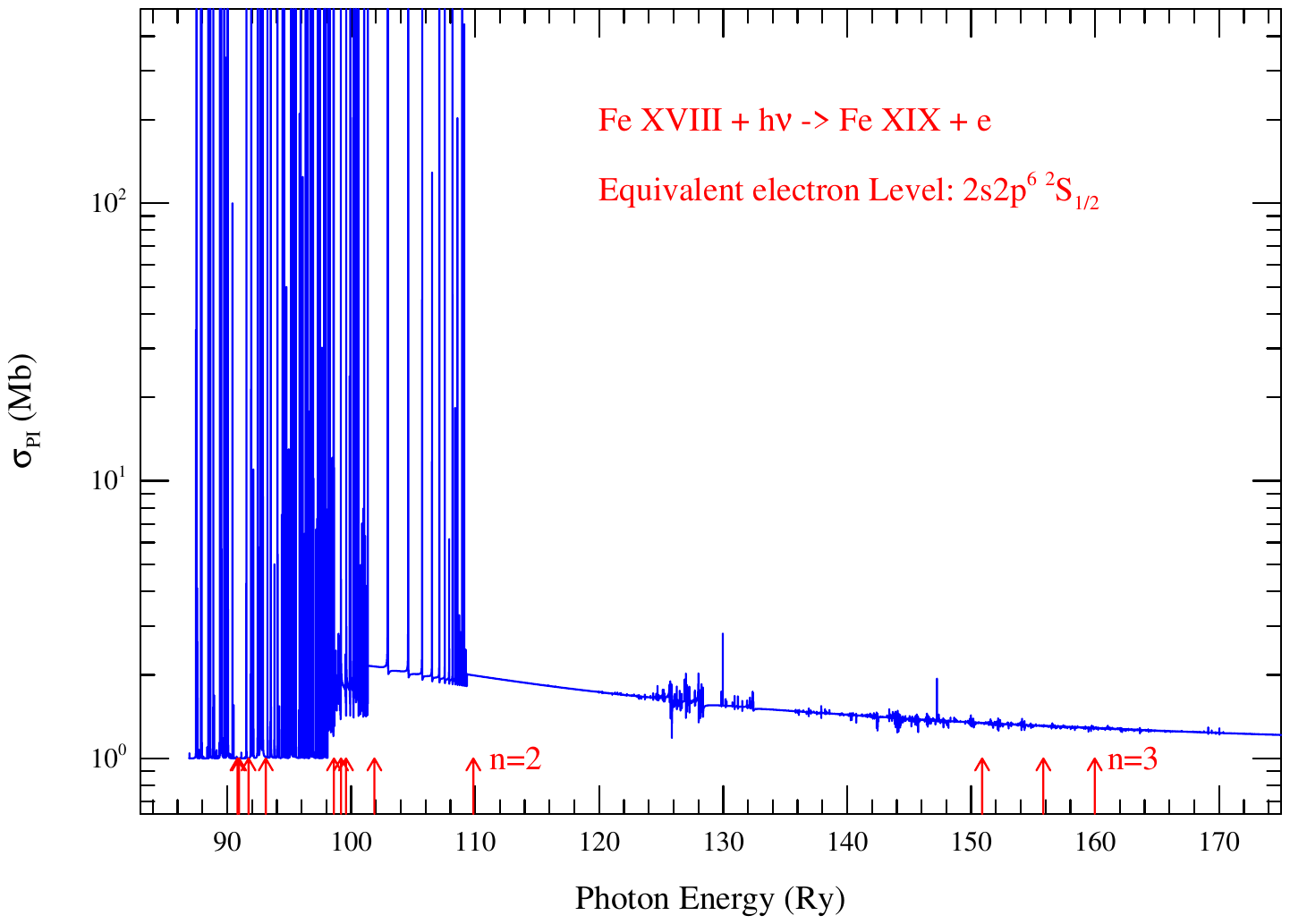}
 \caption{$\sigma_{PI}$ of the equivalent electron state
$2s2p^6(^2S_{1/2})$ of \fexviii. Arrows at the bottom indicate
excitation thresholds to which Rydberg series of resonances converge. 
Arrows at n=2 and n=3 indicate the
highest core ion excitation energy for the respective shell. Features
show formation of very strong series of \en = 2 resonances; higher ones 
are much weaker. 
\label{fig:f4}
}
\end{figure}

\begin{figure}
\hskip -0.35in
  \includegraphics[height=4.0in,width=5.0in]{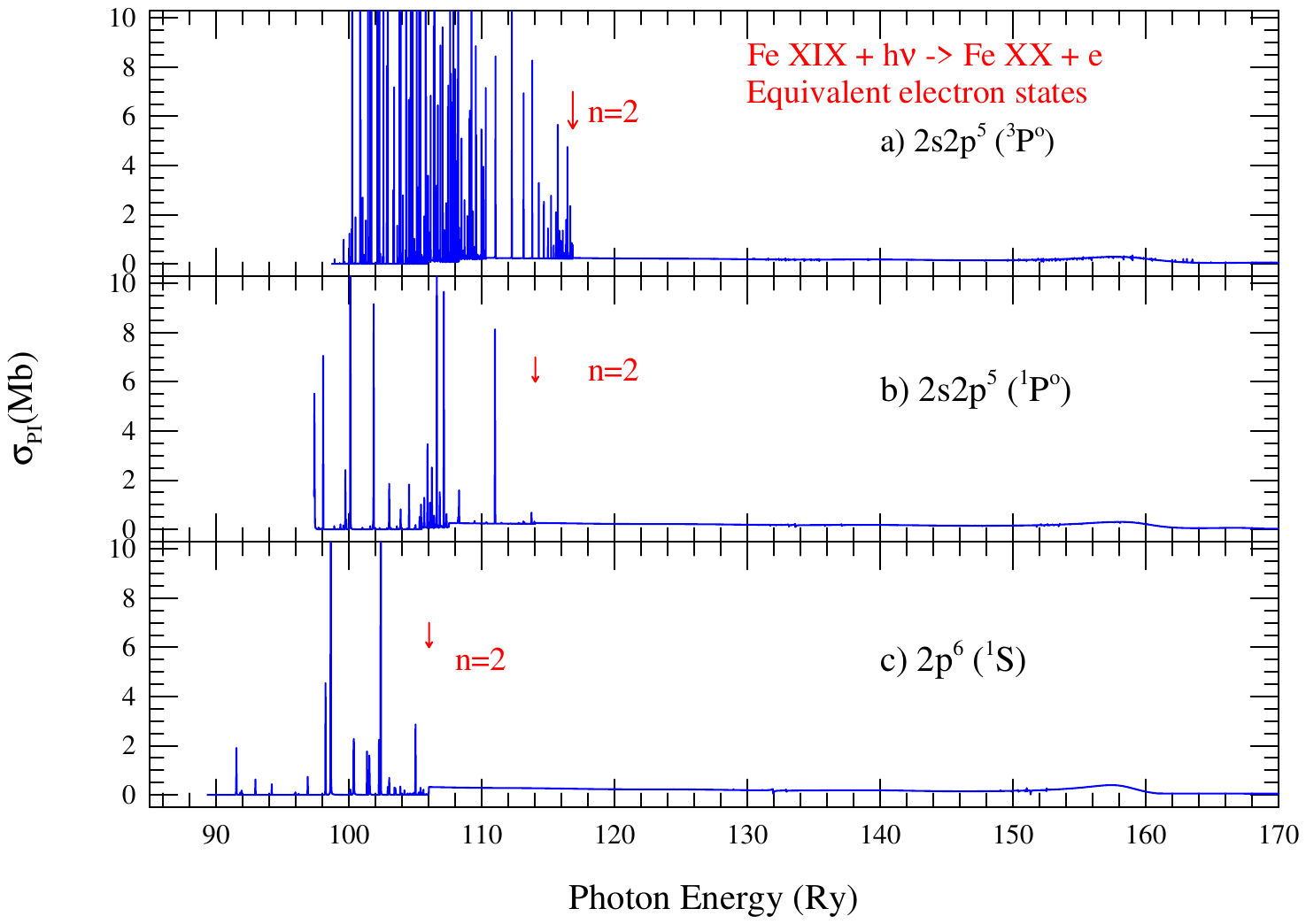}
 \caption{$\sigma_{PI}$ of the equivalent electron states a) $2s2p^5(^3P^o)$,
b) $2s2p^5(^1P^o)$, and c) $2p^6(^1S)$  of Fe~XIX. Arrows at n=2 indicate 
the highest core ion excitation energy threshold for the shell and
appearance of resonances below it. Features
show formation of very strong dense series of resonances for 
$2s2p^5(^3P^o)$ for stronger coupling while it they are less for the 
singlets due to lack of coupling couplings to the states. 
\label{fig:f5}
}
\end{figure}

For the present three ions, there is no equivalent electron state 
for \fexvii, one for \fexviii~ and three for \fexix. Photoionization
cross sections ($\sigma_{PI}$) of these levels for \fexviii~ and \fexix~
are shown in Figures~\ref{fig:f4} and \ref{fig:f5}. 
Figure~\ref{fig:f4} presents $\sigma_{PI}$ 
of the equivalent electron state $2s2p^6(^2S_{1/2})$ of \fexviii. The
arrows point to positions of various core ion thresholds where Rydberg
series of resonances converge and enhance the background. It may 
be noted that closely-spaced  Rydberg resonances are highly-peaked 
and strong only for n=2 core ion thresholds. Higher ones
are very weak resonances merging with the background.

Figure~\ref{fig:f5} presents $\sigma_{PI}$ of the three equivalent 
electron states 
of Fe~XIX: a) $2s2p^5(^3P^o)$, b) $2s2p^5(^1P^o)$, and c) $2p^6(^1S)$.
Features are similar to those of Figure~\ref{fig:f4}, with almost no
resonance structures beyond \en=2 threshold
excitations. However, note that 
$\sigma_{PI}($$2s2p^5(^3P^o))$ in panel a) shows closely-spaced 
strong resonances, 
those are sparse for the other two levels in panels b) and c). 
The reason for the latter 
is lack of the channel couplings with the singlet
states $2s2p^5(^1P^o)$ and $2p^6(^1S)$.

\subsubsection{Photoionization of low-lying excited levels:}

\begin{figure}
\hskip -0.35in
 \includegraphics[height=3.0in,width=5.0in]{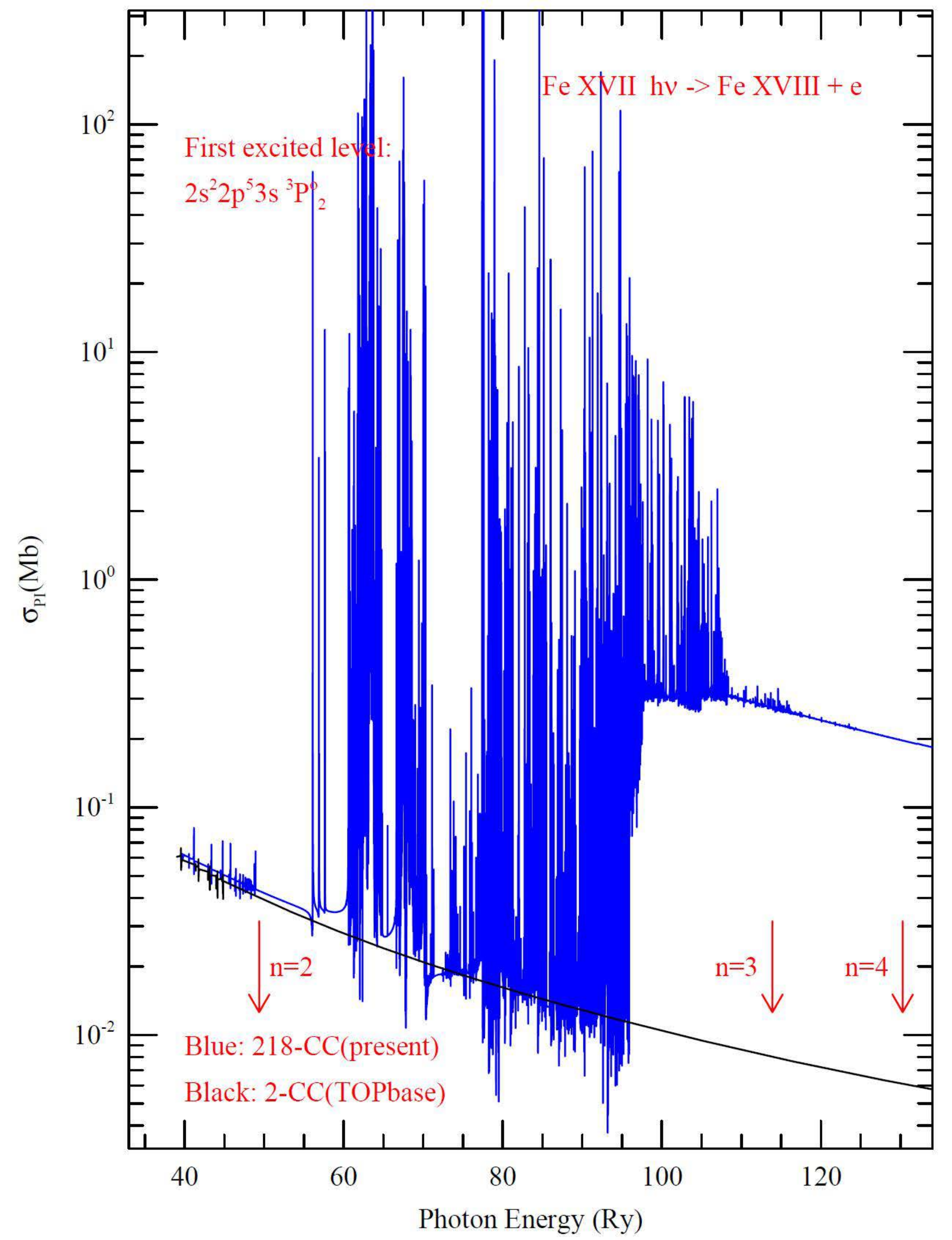}
 \caption{
$\sigma_{PI}$ of the first single valence electron excited level of 
$2s^22p^53s(^3p^o_2)$ of \fexvii, demonstrating impact of the \en=3 
complex of resonances forming strong resonance features that are 
several orders of magnitude higher than those due to the \en=2 complex 
and background enhancement near the \en=3 threshold; however, beyond 
the \en=3 resonances are very weak.  \label{fig:f6}
}
 \end{figure}

Photoionization features change dramatically for single-electron excited 
states in comparison to those of the ground and equivalent electron states. 
Core excitations to high lying levels beyond the 
ground \en-complex exhibits enriched variations. 
For the three Fe ions, considerable impact can be seen in $\sigma_{I}$ of
excited states in forming strong 
resonances and enhancing the background beyond the ground n-complex 
(n=2), even for the first excited level illustrated in Figure~\ref{fig:f6}.

Figure~\ref{fig:f6} presents $\sigma_{PI}$ of the first single valence 
electron excited level $2s^22p^53s(^3p^o_2)$ of \fexvii; 
blue represents this work 
and black the OP data obtained by M. P. Scott (unpublished) available 
in TOPbase \cite{top}. Regions of 
resonant features belonging to core excitation to \en=2, \en=3 and \en=4 
complexes are marked by arrows which point to 
energies of the highest excited core of the respective \en-complex; 
those associated with \en=2 are very weak and the background
cross section is decreasing. However, above ~$\sim$ 57 Ry strong
resonance structures appear and dominate until \en=3 thresholds
around $\sim$97 Ry, where 
the background rises by more than on order of
magnitude. Beyond \en=3, resonances become weak and
merge with the smooth background which decreases smoothly.
The strength of the \en=3 resonance complex, relative to \en=2,
indicates high photoabsorption and enhanced background 
missing in OP data \cite{top}.

\begin{figure}
\hskip -0.35in
  \includegraphics[height=3.0in,width=5.0in]{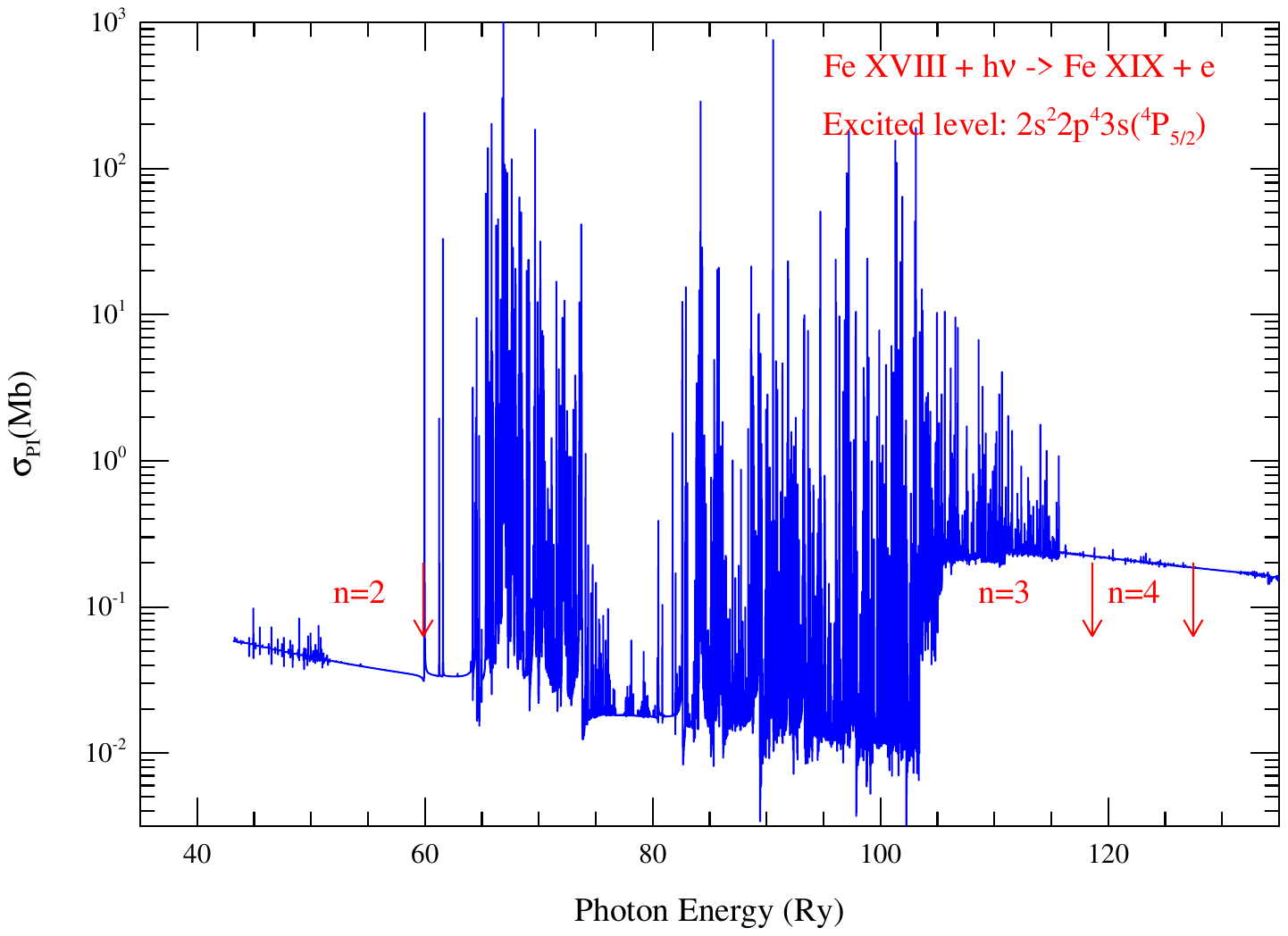}
 \caption{ $\sigma_{PI}$ of the first single electron excited level 
$2s^22p4^43s(^4P_{5/2}$ level of \fexviii~ demonstrating
 core excitations of the \en=3 complex similar to that of \fexvii~
in Figure~\ref{fig:f6}.
\label{fig:f7}
}
 \end{figure}

Figure~\ref{fig:f7} presents $\sigma_{PI}$ of the first single valence 
electron excited level $2s^22p^43s(^4Po_{5/2})$ of \fexviii~ where the 
arrows point to the energy positions of the highest excited core of 
the respective shells \en=2,3,4. Although it has one less electron,
F-like instead of Ne-like \fexvii, the features are similar indicating
characteristics of the level:
large enhancement due to \en=3 complex of resonances relative to \en=2.
Couplings to the \en=4 complex are very weak and the background
decreases with energy. Overall, between $\sim$60-102 Ry 
resonant excitations raise the background by more than an order of 
magnitude, and merge into the background above the \en=3 thresholds.

\begin{figure}
\hskip -0.35in
  \includegraphics[height=3.0in,width=5.0in]{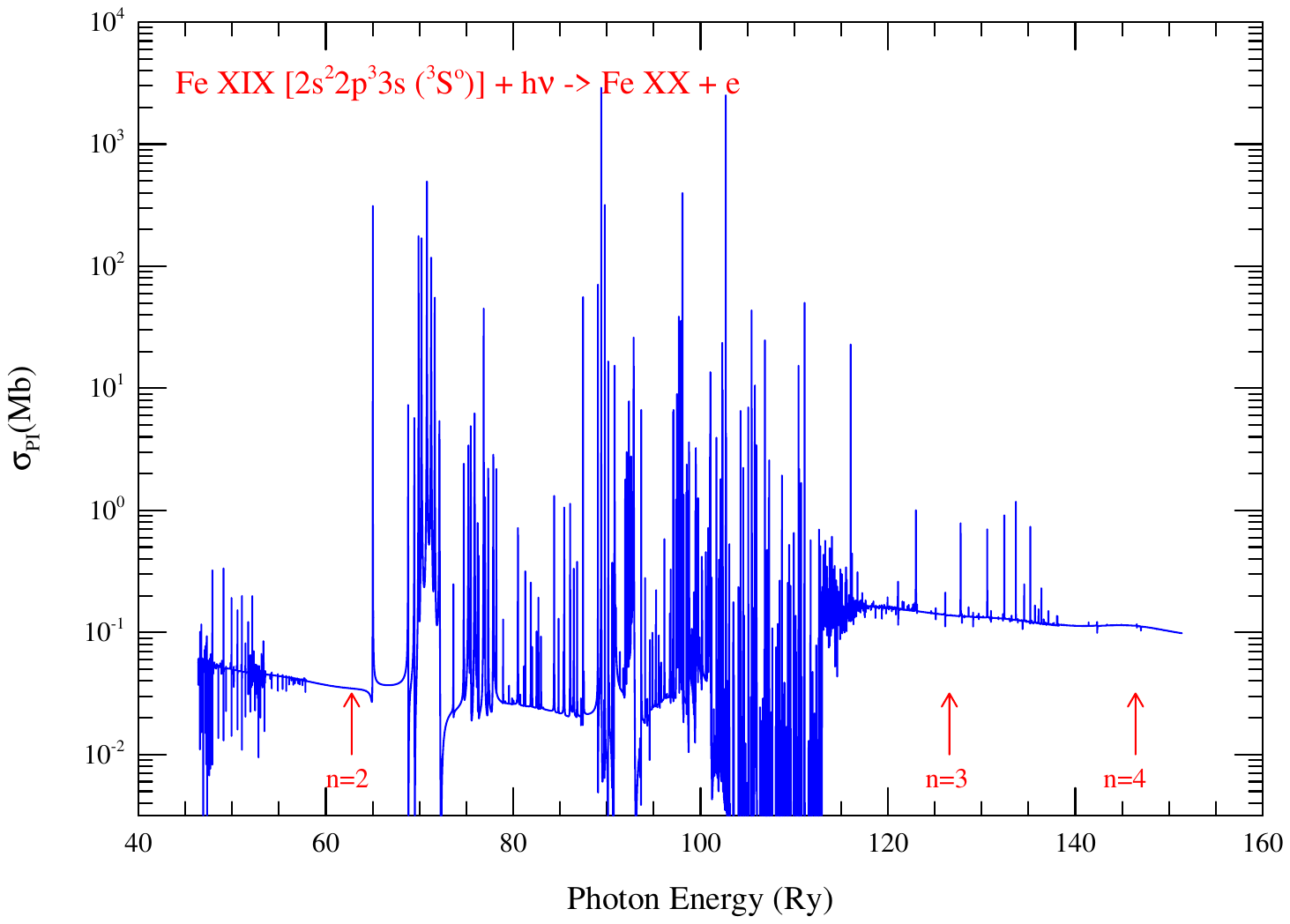}
 \caption{ $\sigma_{PI}$ of the first excited level of 
$2s^22p^33s ^3S^o$ of \fexix, as for Figures~\ref{fig:f6} and \ref{fig:f7}.
\label{fig:f8}
}
 \end{figure}

Figure~\ref{fig:f8} presents $\sigma_{PI}$ of the first single electron 
excited state,
$2s^22p^33s ^3S^o$ of \fexix~ demonstrating relative magnitudes of
resonant complexes due to \en=2 and \en=3 thresholds,
However, similar to \fexvii~ and \fexviii~
the impact is negligible for \en=4 complex. {\it Therefore, it may be
concluded that resonance structures due to core excitations
 beyond \en=4 for all three Fe ions have converged.}

\subsubsection{Seaton PEC resonances:}

\begin{figure}
\hskip -0.35in
  \includegraphics[height=3.5in,width=5.5in]{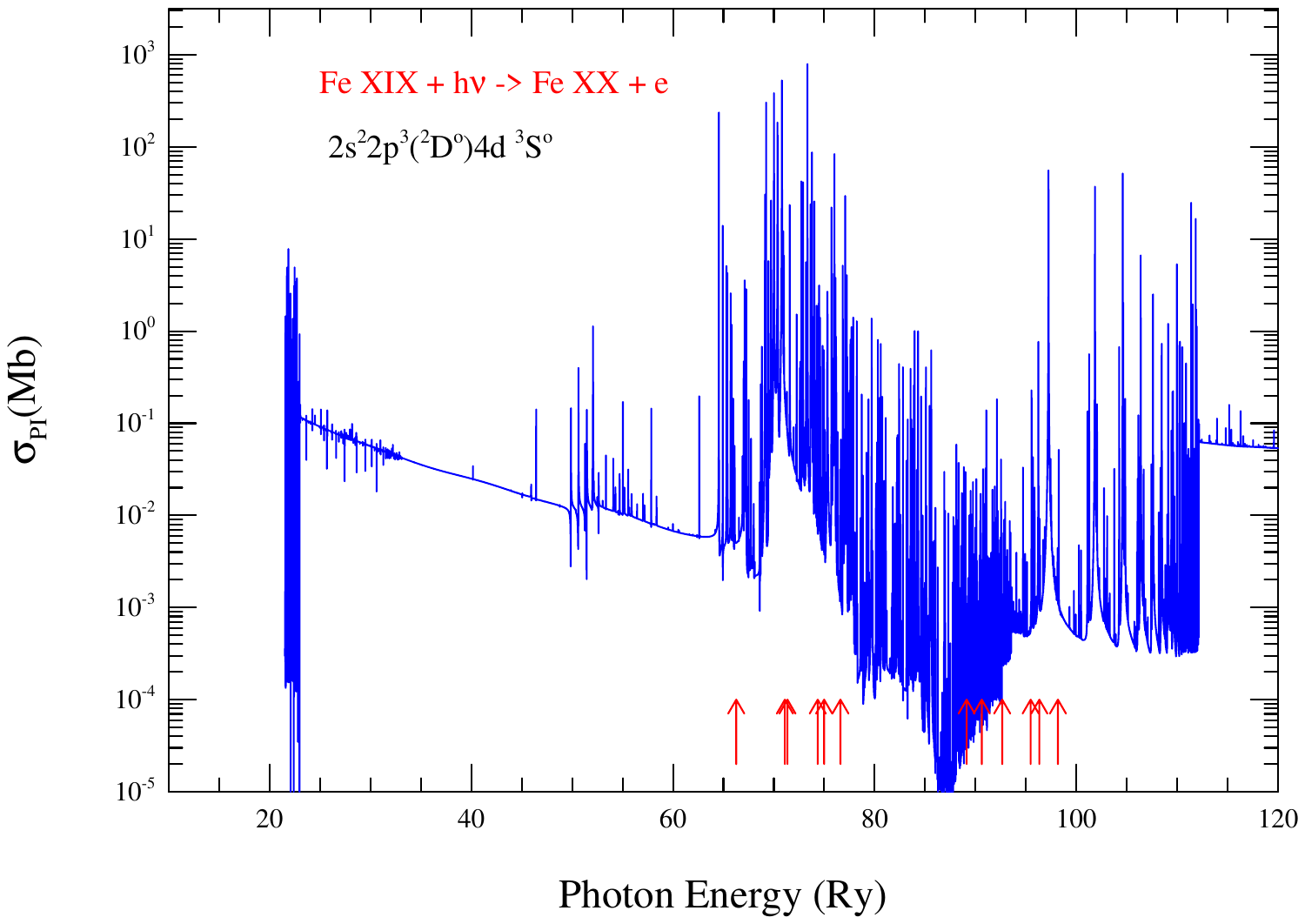}
 \caption{ Characteristic features of a high-lying excited state,
$2s^2p^3(^2D^o)4d(^3S^o)$, of \fexix. Photoionization of such levels 
is typically dominated
by wide Seaton PEC resonances, as seen at energy positions pointed by red 
arrows. Interference of Seaton and Rydberg resonances can impact the
background over a wide energy range up to and above 100 eV, and 
enhancement by more than an order of magnitude above the background. 
\label{fig:f9}
}
 \end{figure}

In addition to dense and strong Rydberg series of resonances, the high 
lying excited states typically show dominance by huge resonances 
due to photo-excitation-of core via strong dipole transitions 
over extended energy regions \cite{ys87,aas11,snn18}. Figure~\ref{fig:f9}
presents $\sigma_{PI}$ of the
high-lying single valence electron excited level $2s^2p^3(^2D^o)4d(^3S^o)$, 
of \fexix. Generally, Seaton PEC resonances 
are formed as the core ion is excited from the ground state through
a dipole allowed transition, while the outer electron remains a
spectator as it photoionizes \cite{nb5}. 
The interference of Seaton and Rydberg resonances usually form a
combined resonance feature over a wide energy range, as
seen around $\sim$70 Ry and $\sim$95 Ry (red arrows) in Figure~\ref{fig:f9}.
Since the transition energies remain the same irrespective of
ionization thresholds, Seaton resonances appear at the same photon
energies in $\sigma_{PI}$ of all excited levels, but not in $\sigma_{PI}$ 
of equivalent electron states shown in Figure~\ref{fig:f5}.

\begin{figure}
\hskip -0.35in
  \includegraphics[height=5.0in,width=5.5in]{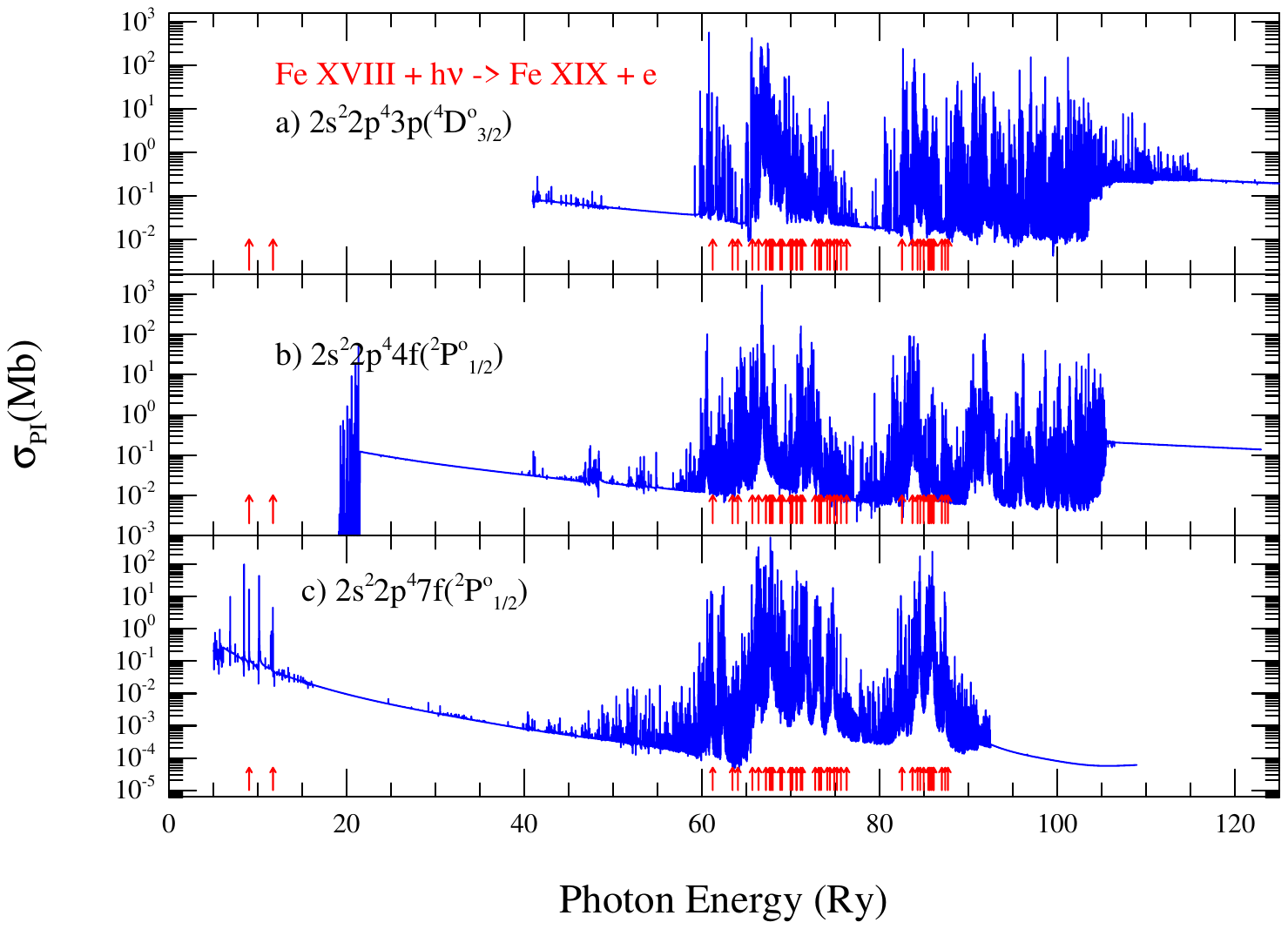}
 \caption{Progression of chararcterstic features 
with excitation level:
a) $2s^2p^43p(^4D^o_{3/2})$, b) $2s^22p^44f(^2P^o_{1/2})$ and 
c) $2s^22p^47f(^2P^o_{1/2})$ of Fe~XVIII. Seaton resonances appear at the exact
energies of core transitions via dipole allowed transitions, and become
more prominent in $\sigma_{PI}$ with higher level of \en-value of the 
spectator electron.
\label{fig:f10}
}
 \end{figure}

Figure~\ref{fig:f10} illustrates other characteristics of Seaton resonances,
 the progressive behaviour and commonality in
positions at the same PEC energies. Figure~\ref{fig:f10} shows 
$\sigma_{PI}$ of three excited levels with different ionization energies:
a) $2s^2p^43p(^4D^o_{3/2})$ at $\sim$41 Ry, b) $2s^22p^44f(^2P^o_{1/2})$ 
at $\sim$19.07 Ry and c) $2s^22p^47f(^2P^o_{1/2})$ at $\sim$5.06 Ry
of Fe~XVIII.  The first two resonances appear around 10 Ry, but
seen only for the third excited state (panel c) since the ionization 
threshold for the level is lower than for these PEC excitations.
Other Seaton resonances are at higher energies and appear in
$\sigma_{PI}$ of all levels exactly at the same transition energies.  
The general feature is that Seaton resonances become more prominent with
exciation level, from a) to c) in 
Figure~\ref{fig:f10}.  
Finally, we also find that Seaton resonances with $\Delta n$=1 are 
stronger than than those with $\Delta n$=0, but weaker for higher core
transition energies $\Delta n$. 

Finally, these PEC phenomena should be detectable experimentally owing to their
extended energy ranges. By virtue of their immense magnitude
and extent, Seaton PEC resonances are the largest contributor to
bound-free opacity.

\subsubsection{Convergence of resonance series:}

One crucial outcome of having a vary large wavefunction expansion which
includes many core ion excitations is the convergence of highly-peaked
prominent resonant features with increasing principle quantum number n.
Exceptions are the $\sigma_{PI}$ of the ground level and equivalent 
electron levels for which resonances become weaker and start to converge 
to the background beyond \en=2 complex. However, altogether the
ground and equivalent
electron states are relatively few in number compared to hundreds to
over a thousand other excited bound states of each ion where AI
resonances dominate.
With increasing \en, the excitation probability of the core ion first
increases and then starts to decrease and weaker channel couplings
merging on to the background.
In $\sigma_{PI}$ presented in Figures~\ref{fig:f3}-\ref{fig:f10} 
for the three Fe ions, we see various progressions with \en=2, 3 and 4.
The n=4 AI
resonance structures are reduced considerably and show the trend towards
convergence.
Past computations of $\sigma_{PI}$ either did not consider resonances
beyond the \en-complex of the ground configuration, or 
prominent Rydberg and Seaton resonances and their convergence.

\section{Conclusion}

We have presented detailed albeit limited sets of 
photoionization cross sections of three Fe ions \fexvii, \fexviii~ and
\fexix~ that are the dominant iron opacity source in the
solar interior at the boundary of the radiative and convection zones
(paper RMOP.I). We explore features 
in the high energy region that include core ion excitations in 
hundreds of levels of these three ions. These features were not
heretofore studied primarily owing to presumption of
of weak couplings of interacting channels and AI resonance
interference effects, as well as practical computational limits. 
The present study reveals characteristic features
of photoionization based on 
the type of initial bound states and convergence criteria of AI resonant
phenomena, and relativistic fine structure effects not produced in LS
coupling.
Application of these data are expected to provide    
high-precision plasma opacity in stellar interiors for these ions.

{\bf Data availability}

All atomic data for energies, radiative transitions, and collisional
excitations will be available online at the NORAD-Atomic-Data database
at the Ohio State University at:
https://norad.astronomy.osu.edu/

\subsection{Acknowledgments}

We acknowledge the support of the Ohio Supercomputer Center for
extensive computational resources for the present work.
Computations were carried out over several years on 
high performance computer clusters at 
the Ohio Supercomputer Center (OSC) and on the UNITY cluster at the Ohio State 
University.

\pagebreak


\begin{thebibliography}{}
\bibitem{nb1} Computations also faced numerical challenges for complex atomic
systems, such as for photoionization of
Fe~II (\cite{np94} the largest calculation at that time), and
difficulties with supercomputer systems 
that also limited the scope of atomic calculations.
\bibitem{nb2} Fine structure transitions were obtained from 
LS multiplets of various 
ions through algebraic transformation and using observed energies by
the National Institute of Standards and Technology (NIST) \cite{nist} and
Nahar (e.g. \cite{snn95})
\bibitem{nb4} The equation-of-state discussed in RMOP.I determines 
level populations at a given temperature and density. It is found that
only a few percent of total ionic populations may be in the ground state
in stellar interiors where thd Fe ions considered herein in are
prevalent.
\bibitem{nb5} This is the inverse of the dielectronic reombination process when
the photoexcited core radiatively decays when a free electron recombines
via PEC resonances.
\end{thebibliography}
\end{document}